# The Role of Urban Designers in the Era of AIGC:
# An Experimental Study Based on Public Participation

Di Mo[1], Keyi Liu[1], Qi Tian[1], Dengyun Li[1], Liyan Xu[1*], Junyan Ye[1*],

1College of Architecture and Landscape, Peking University

* Co-Corresponding Author

modi2022@stu.pku.edu.cn, ky.liu@stu.pku.edu.cn, 2301213461@stu.pku.edu.cn, 2201221756@stu.pku.edu.cn,

xuliyan@pku.edu.cn, yejunyan@pku.edu.cn

**Abstract**

This study explores the application of Artificial Intelligence Generated Content (AIGC) technology in urban planning and design, with a particular focus on its impact on placemaking and public participation. By utilizing natural language processing and image generation models such as Stable Diffusion, AIGC enables efficient transformation from textual descriptions to visual representations, advancing the visualization of urban spatial experiences. The research examines the evolving role of designers in participatory planning processes, specifically how AIGC facilitates their transition from traditional creators to collaborators and facilitators, and the implications of this shift on the effectiveness of public engagement. Through experimental evaluation, the study assesses the design quality of urban pocket gardens generated under varying levels of designer involvement, analyzing the influence of designers on the aesthetic quality and contextual relevance of AIGC outputs. The findings reveal that designers significantly improve the quality of AIGC-generated designs by providing guidance and structural frameworks, highlighting the substantial potential of human-AI collaboration in urban design. This research offers valuable insights into future collaborative approaches between planners and AIGC technologies, aiming to integrate technological advancements with professional practice to foster sustainable urban development.

## Introduction

Artificial Intelligence (AI) refers to technologies and systems that enable machines to simulate human intelligence behaviors (Mehrabanian, 2023). Among these, Artificial Intelligence Generated Content (AIGC) leverages AI algorithms to assist or automate the creation of media content such as text, images, audio, and video. AIGC is transforming urban planning and design, a field traditionally reliant on human creativity and expertise, by introducing AI's capabilities for intelligent analysis and automated generation. AIGC's evolution is reshaping planning tools and processes, prompting critical questions: Can AIGC serve as both an efficient tool and an innovative approach? How does it influence designers' roles, and how can they adapt to this technological shift to enhance professional practices?

Urban design research in the 20th century has undergone a paradigm shift from a focus on "space" to an emphasis on "place," integrating intangible aspects like human activity with tangible urban elements (Seamon & Sowers, 2008). The formation of place is a synthesis of interactions among physical space, events, and meanings (Najafi & Shariff, 2011), requiring a holistic understanding of urban environments. Placemaking involves a comprehensive approach to material space (focused on architecture and public spaces), lived space (centered on everyday life and local traditions), imagined space (rooted in experiential and memory elements), and the metaphorical space generated by the interplay of these dimensions.

Advancements in AIGC technology resonate with place theory across spatial, cultural, and semantic dimensions. AIGC can translate textual descriptions into visual representations using advanced models like Stable Diffusion. These models analyze emotional tones, spatial cues, and behavioral descriptions in the text, creating visual outputs aligned with place theory (Radford et al., 2021; Wartmann & Purves, 2018). Training data are critical, allowing AIGC to map textual semantics to visual symbols, such as associating "Eastern courtyard" with Chinese tiled roofs or "Nordic village" with wooden structures (Dolhopolov et al., 2024). Techniques like U-Net architectures in diffusion models refine these outputs from coarse to fine detail (Rombach et al., 2022). By refining textual input analysis, leveraging multi-level training data learning, and precisely balancing overall and localized relationships during generation, AIGC technology can partially reconstruct the material and symbolic meanings of place. The challenges posed by AIGC technology to urban planning particularly influence participatory planning, a process emphasizing public involvement to better reflect community needs and aspirations (Lane, 2005). By transforming the working methods of planners, lowering the barriers to public engagement, and potentially reshaping the relationship between citizens and the planning process, AIGC brings significant changes that could directly threaten the professional role of designers. Originating in the 1960s and 1970s as a response to the limitations of top-down planning methods, participatory planning relies on public feedback and designer expertise to translate abstract ideas into actionable designs (Forester, 1982; Healey, 1992).

Traditional public participation in planning relied on methods like interviews, surveys, and real-time feedback to collect textual input. However, these approaches often lacked intuitive visual communication, particularly for urban environmental issues. Designers played a key role in translating abstract public ideas into concrete visual proposals. The rise of AIGC technology, particularly text-to-image AI models based on diffusion frameworks such as Stable Diffusion, MidJourney, and DALL-E 2, has significantly bridged the gap between textual descriptions and psychological imagery. These tools allow images to be generated and modified based on text input and even enable model fine-tuning to produce consistent stylistic outputs (Borji, 2022). By providing a more intuitive means for the public to express their perspectives and expectations for urban environments (Kim et al., 2022), AIGC technology democratizes design expression. Additionally, integrating AIGC with human-computer interaction (HCI) technology offers a more intelligent and natural interactive experience (Aneja et al., 2021). These advancements empower non-experts to articulate their design ideas using simple textual inputs, bypassing the traditional role of designers. This shift risks marginalizing designers, placing them in a precarious position within the planning process.

Given this context, it is critical to investigate how AIGC technology influences the role of designers and redefines their responsibilities and value in urban planning. The study evaluates the quality of site redesign schemes generated by AIGC tools with varying levels of designer involvement during public participation processes. Our contributions can be summarized as follows:

- Fine-tune a model tailored for street garden scenes, enabling Stable Diffusion to generate design schemes.
- Recruit 160 participants and analyze the aesthetic and alignment scores of AIGC design outputs under different designer involvement levels.
- Explore the boundaries of AIGC tools in the field of urban design and ways of collaborating with human designers.

## Research Methods

This study employs advanced diffusion models in the field of artificial intelligence image generation to produce street views that meet urban planning professional standards. The methodology integrates design knowledge and spatial relationships into the model through fine-tuning and soft inpainting techniques to ensure that generated outputs align with planning requirements. The system comprises three core modules: algorithm architecture, knowledge graph construction, and a public participation platform.

### Diffusion Model Architecture

The study adopts an advanced diffusion model combining soft inpainting and the latent diffusion-based Stable Diffusion model to achieve semantically guided urban street view generation (Fig. 1).

- **Soft Inpainting:** Soft inpainting provides a flexible mechanism for localized image modifications, seamlessly blending the original street view with reconstructed elements to ensure a natural and cohesive overall effect. By smoothly filling missing regions, the technique enables seamless connections between the original and newly generated components, ensuring visual realism. In street view generation and editing, soft inpainting retains details of elements such as buildings and roads while naturally integrating masked areas, producing high-quality, coherent urban street images with enhanced visual fidelity.

- **Latent Diffusion-Based Stable Diffusion Model:** The Stable Diffusion model generates and modifies high-quality urban street view images. By integrating region-specific information from soft inpainting, the model allows user-defined areas to be modified according to provided textual descriptions. The model can comprehend and generate complex visual effects effectively, ensuring that outputs align with user-defined design requirements.

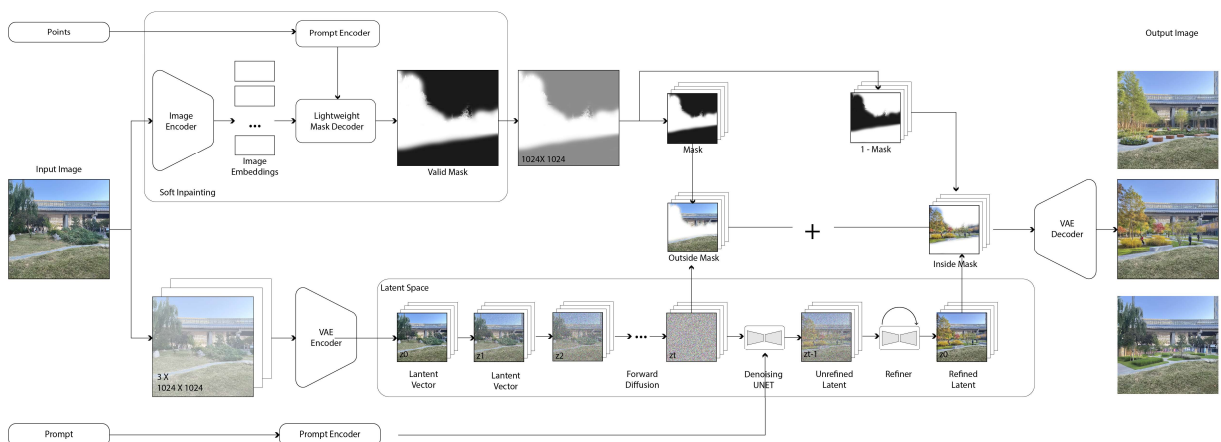

Figure 1 Model training progress.

## Integration of Professional Knowledge

To ensure the model reflects the expertise of urban planners, two knowledge graphs were developed: one for design elements and another for spatial relationships. These graphs were translated into tag sets used for model training, fine-tuning a model tailored for street gardens, and embedding professional knowledge into the generation process (Fig. 2).

- **Design Elements:** The design elements knowledge graph includes functionality, design style, overall ambiance, plant design, materials, and other design aspects. Tags derived from this graph guide the model in understanding design requirements, enabling the generation of street views that meet expectations. For example, functional tags can direct the intended usage scenarios of a street view, while style tags influence the overall aesthetic.
- **Spatial Relationships:** The spatial relationships knowledge graph encompasses scale and perspective relationships, guiding the model to convey realistic spatial proportions during generation. These tags ensure that generated street views exhibit accurate visual effects, enhancing aesthetic quality and consistency.

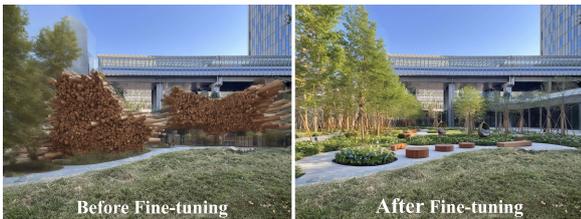

Figure 2 Generated outputs before and after fine-tuning.

## Public Participation Platform

A public participation platform was developed using the inpainting functionality of the Stable Diffusion WebUI. This platform simplifies the process of involving the public in urban street design and transformation by lowering technical barriers.

- **Interactive Inpainting WebUI:** The WebUI enables interactive, localized modifications of street views. Users can freely select parts of urban street images for modification, and the interface integrates seamlessly with SAM segmentation models and Stable Diffusion, allowing real-time updates and previews of changes. This interactive approach reduces cognitive load, enabling users to engage in complex urban design processes through simple operations.
- **Ease of Use:** The WebUI is compatible with smartphones and tablets, allowing users to participate in urban environment annotation and transformation anytime, anywhere. The localized inpainting feature simplifies user interactions, enabling intuitive street view modifications through straightforward clicks and selections. This approach facilitates a seamless "what-you-see-is-what-you-get" experience for street view design. This interface makes people collaboratively shape urban environments easier.

## Research Design

This research evaluates how AIGC facilitates public engagement and whether outputs generated without designer involvement meet expected standards. A key focus is on the collaboration between designers and AIGC tools during participatory planning. The study examines how varying levels of designer involvement—ranging from "no guidance" to "moderate" and "intensive guidance"—influence the quality and creativity of design outcomes. The primary objective is to assess satisfaction and alignment of urban pocket garden designs generated under different designer involvement levels. Additionally, the study compares design quality and acceptance across groups with and without design backgrounds, offering insights into the effectiveness of AIGC in public participation (Fig. 3).

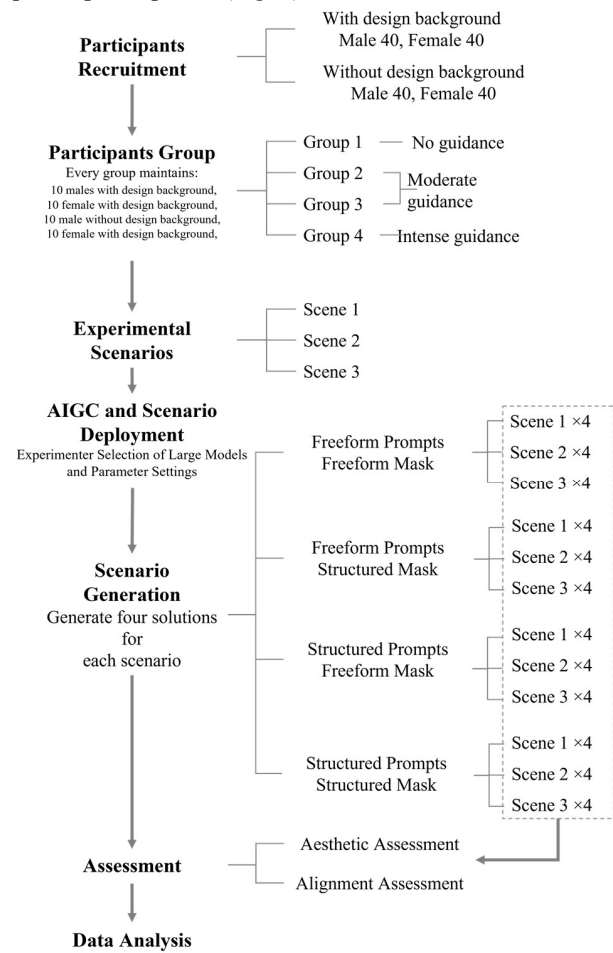

Figure 3 Experimental Procedure.

## Hypotheses

### Hypothesis 1

The level of freedom granted to users during the AIGC generation process significantly affects the quality of the generated images.

- Combinations of freeform sketches and unrestricted keywords result in the lowest quality outcomes.
- Combinations of pre-defined sketches and structured prompts produce the highest quality outcomes.

**Hypothesis 2**

The level of designer involvement significantly impacts the quality and goal alignment of the generated designs.
- Results generated by groups with designer involvement are significantly higher in quality compared to those without designer involvement.
- Designs generated under intensive guidance conditions exhibit the highest quality.

**Hypothesis 3**

Professional background has a significant impact on the quality of the generated designs.
- Participants with a design background produce significantly higher-quality results than those without a design background.

## Experimental Variables

### Experimental Scenarios

Three typical urban pocket garden scenarios were selected for this study, each characterized by distinct spatial forms, internal terrain, and landscape elements. These scenarios represent diverse and representative features of urban roadside green spaces, minimizing potential interference from specific green space types (Fig. 4).

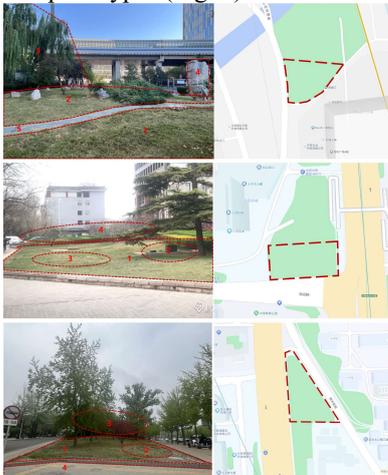

Figure 4 Locations and Structured Mask of Scenes.

### Methods for Selecting Modification Areas

- **Freeform Mask:** Participants are allowed to freely select areas within the images for modification. This approach enables observation of the breadth of user creativity and the unpredictability of the generated results.
- **Structured Mask**: Researchers predefine the modifiable areas, restricting participants' selection. This method focuses on key areas and facilitates to comparison of the effects of different prompts on the same area.

### Methods for Providing Prompts

- **Freeform Prompts:** Participants are permitted to freely describe their design intentions. This approach captures personalized creativity and inspiration but carries the risk of deviations in the generated results due to inaccuracies in expression.
- **Structured Prompts:** Designers provide a predefined set of prompt modules, from which participants must select those that match their intentions. Structured prompts enhance the predictability and consistency of the generated results.

## Experimental Group Setup

### Participants Groups

Healthy participants aged 21–30 are recruited, with psychological evaluations conducted to exclude those with mental disorders, excessive anxiety, or unstable stress levels. The study avoids recruiting the researchers' subordinates. In addition to ensuring equal numbers of genders, participants were evenly divided into the following two groups.

- **With Design Background:** Participants possess a professional background in spatial design fields, such as architecture or landscape design, and demonstrate a foundational knowledge of design principles and skills.
- **Without Design Background:** General public participants with no professional design experience. Their interaction with AIGC is used to evaluate the effectiveness of public participation without designer assistance.

According to gender and professional background, all participants were divided into four groups: men with a design background, women with a design background, men without a design background, and women without a design background. Sample size calculations using Gpower 3.1 require 18 participants per group (Power = 0.95, α = 0.05). To ensure validity, 20 participants per group are recruited, totaling 160 participants.

### Levels of Designer Involvement

- **No Guidance:** AIGC generates designs solely based on user input without any designer intervention. This setup simulates a scenario where designers are entirely marginalized, allowing the high freedom of participants.
- **Moderate Guidance:** Designers intervene by providing structured prompts or suggesting pre-defined mask areas to modify, correspondingly the participants are under a middle degree of freedom.
- **Intensive Guidance:** Designers offer detailed design guidance, including predefined mask areas and precise structured prompts, and now the participants are at a low degree of freedom.

Participants are randomly assigned to experimental conditions using random sampling tools, ensuring equal numbers per condition.

## Evaluation of Generated Results

- **Aesthetic Assessment:** Participants rate the aesthetic quality of the generated outcomes on a scale of 1 to 10.

- **Alignment Assessment:** Participants evaluate the alignment of the generated outcomes with the initial design intent, using a scale of 1 to 10.

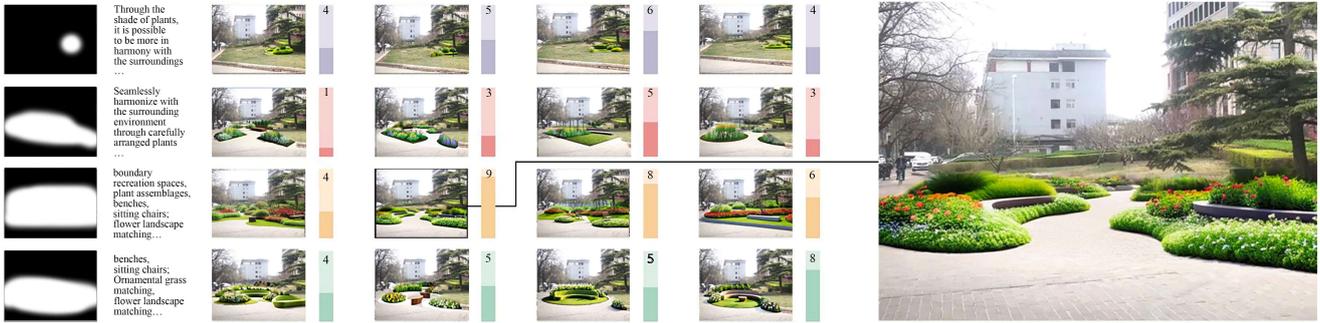

Figure 5 Partial experiment results.

## Data Processing and Analysis

### Data collection and preprocessing

Stratified sampling by age and gender ensures balanced recruitment for design and non-design groups. Data collection is anonymized, and a single-blind design minimizes participant biases, while administrators oversee precise execution. Participants needed to evaluate three sets of scene pictures generated by AI tools, and each scene contained four generated schemes. During the experiment, after the aesthetic and alignment scores of each participant were recorded, a semi-structured interview was conducted to collect their textual feedback.

Calculated the mean aesthetic and alignment scores of four schemes to represent each scene's final scores. And calculated the mean scores of three scenes to represent each participant's final scores. Under the freeform prompts guidance mode, the scores of modified pictures were grouped separately for subsequent analysis (Fig. 5).

### Result Analysis with different degrees of freedom

#### Comparison of different levels of freedom

In this paragraph, the results of the experimental groups were compared according to the division of different degrees of freedom. As for aesthetic scores, the sequence of its median was "medium degree of freedom> low degree of freedom> high degree of freedom", while the sequence of whose mean scores was "low degrees of freedom> medium degree of freedom> high degree of freedom". As for alignment scores, their median and mean scores also corresponded to the second sequence. The experimental group with a high degree of freedom exhibited higher dispersion compared to other groups, and the dispersion of alignment scores was higher than that of aesthetic scores.

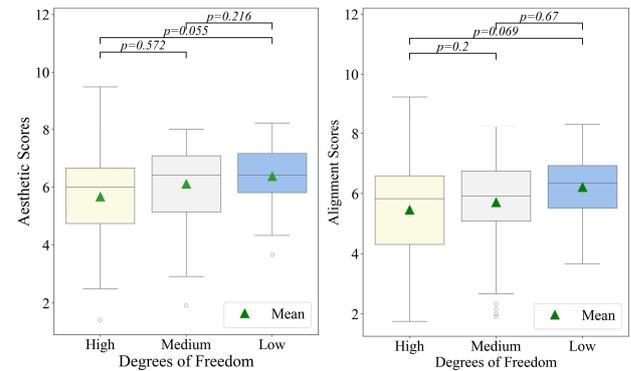

Figure 6 Bloxplot of different degrees of freedom.

ANOVA analysis showed that the differences in the mean scores were marginally significant between various degrees of freedom (the significance level of difference in mean aesthetic scores between various degrees of freedom was 0.065, and that of differences in mean alignment scores was 0.079). Tukey HSD's multiple comparison analysis showed that the differences in mean aesthetic and alignment scores were both marginally significant between the two groups guided by low or high degrees of freedom (significance level was 0.055 and 0.069, respectively), indicating significant score differences between various degrees of freedom groups.

#### Comparison of guiding methods in moderate freedom

Under the medium degree of freedom, the aesthetic and alignment scores (median and mean) of "structured prompts & freeform mask" methods were higher than those of "freeform prompts & structured mask" methods. The former scores were also slightly more dispersed than the latter. The differences in mean scores (both the aesthetic and alignment scores) between the two guiding methods were significant (0.013 and 0.026, respectively), indicating that the guiding mode of structured prompts was more effective under the medium degrees of freedom.

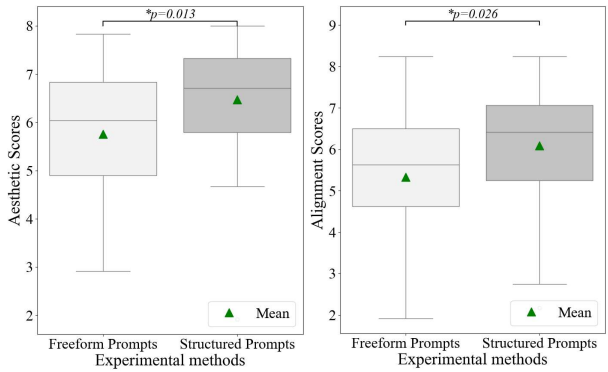

Figure 7 Bloxplot of different guiding methods in medium degrees of freedom.

### Difference analysis before and after polishing of the same guiding method

Under freeform prompts mode ("freeform mask & freeform prompts" and "structured mask & freeform prompts"), the aesthetic and alignment scores (median and mean) after polishing were higher. The dispersion of the aesthetic scores after polishing was also higher, but the alignment scores of "structured mask & freeform prompts" became higher after polishing, while the alignment scores of "structured mask & freeform prompts" were more dispersed without polishing. However, the differences between polishing and no polishing were not statistically significant.

### Analysis with different professional backgrounds

#### Comparison between professional background

Among the results of different experimental groups of professional backgrounds, the aesthetic and alignment scores of nonprofessionals showed a trend of low and high in medium and high degrees of freedom. As for professionals, the sequence of their median aesthetic scores was "medium degree of freedom> low degree of freedom> high degree of freedom", and that of the alignment scores was "low degree of freedom> high degree of freedom> medium degree of freedom". For the different degrees of freedom, the professionals took the high degree of freedom as the most dispersed one, while non-professionals took the medium degree. Otherwise, professionals scored higher under high degrees of freedom, while non-professionals scored higher under medium degrees of freedom.

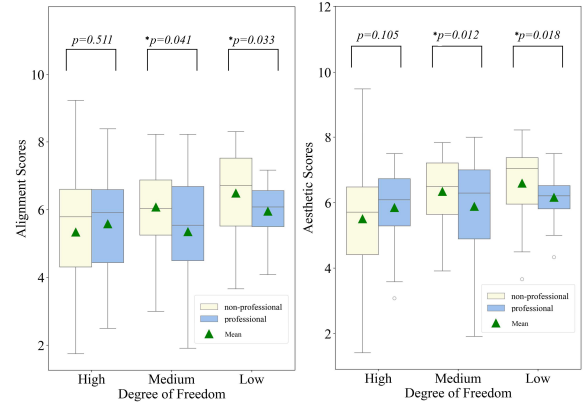

Figure 8 Bloxplot of professional backgrounds in different degrees of freedom.

### Interaction analysis of degrees of freedom and professional background

After comparing the difference of the professional background with description statistics, the interaction effect of the degree of freedom and the professional background was analyzed using the multiple linear regression method. In terms of the main effect of freedom degree, the freedom factor had a significance of 0.064 for aesthetic scores (0.075 for alignment scores), whose η² effect size was 0.035 for aesthetic scores (0.033 for alignment scores). Thus, the degree of freedom could be considered to have a marginal significant main effect on the score result. In terms of the interaction effect of degree of freedom * professional background, the degree of freedom * professional background had a significance of 0.274 for aesthetic scores (0.255 for alignment scores), whose η² effect size is 0.017 (0.018 for alignment scores). The interaction between the degree of freedom and professional background did not reach a statistically significant level.

| Source | Dependent Variable | Type III Sum of Squares | df | Mean Square | F | Sig. | Partial Eta Squared |
|---|---|---|---|---|---|---|---|
| Corrected Model | Aesthetic Scores | 17.307[a] | 5 | 3.461 | 1.918 | .094 | .059 |
|  | Alignment Scores | 26.130[b] | 5 | 5.226 | 2.265 | .051 | .069 |
| Intercept | Aesthetic Scores | 5279.960 | 1 | 5279.960 | 2925.230 | .000 | .950 |
|  | Alignment Scores | 4841.202 | 1 | 4841.202 | 2098.542 | .000 | .932 |
| Guidance Degree | Aesthetic Scores | 10.100 | 2 | 5.050 | 2.798 | .064 | .035 |
|  | Alignment Scores | 12.142 | 2 | 6.071 | 2.632 | .075 | .033 |
| Professional Background | Aesthetic Scores | 1.181 | 1 | 1.181 | .654 | .420 | .004 |
|  | Alignment Scores | 4.170 | 1 | 4.170 | 1.808 | .181 | .012 |
| Guidance Degree* Professional Background | Aesthetic Scores | 4.710 | 2 | 2.355 | 1.305 | .274 | .017 |
|  | Alignment Scores | 6.357 | 2 | 3.178 | 1.378 | .255 | .018 |

|  | | | | |
|---|---|---|---|---|
| Error | Aesthetic Scores | 277.966 | 154 | 1.805 |
|  | Alignment Scores | 355.268 | 154 | 2.307 |
| Total | Aesthetic Scores | 6190.154 | 160 |  |
|  | Alignment Scores | 5721.060 | 160 |  |
| Corrected Total | Aesthetic Scores | 295.273 | 159 |  |
|  | Alignment Scores | 381.398 | 159 |  |

a. R Square = .059 (Adjusted of R Square = .028)
b. R Square = .069 (Adjusted of R Square = .038)

Table 1. Tests of Between-Subjects Effects.

## Extended Analysis

Experimental results reveal the significant impact of designers on the quality of AIGC-generated design outcomes under different levels of intervention, particularly in terms of aesthetic quality and alignment with design intent. Under low freedom conditions, designers effectively controlled the generated content through preset prompts and clearly defined areas for modification, which significantly improved both consistency and alignment (Fig. 6). This approach also reduced the variability in the generated results, ensuring that AI outputs met design goals.

In the medium freedom condition, the scores for the "structured prompt with free drawing" group were significantly higher than those for the "free prompt with structured drawing" group (Fig. 7), indicating that structured guidance enhances the controllability of the generated content, allowing AI to better understand user intent. In this process, the communication skills of designers are particularly important, as clear prompts help convey design concepts effectively, improving the accuracy and consistency of the generated results. As AIGC tools continue to improve, the role of designers as communication bridges will become increasingly crucial, ensuring that professional knowledge is efficiently translated into information understandable by AI, thereby maximizing the quality of collaboration.

Under high freedom conditions, minimal designer guidance led to-- greater variability in content quality, with lower alignment scores and significantly higher scoring variability in the high freedom group (Fig. 6). While greater freedom can foster more creativity, the lack of professional constraints may result in designs that deviate from the intended goals. Thus, the professional judgment of designers remains irreplaceable in ensuring overall spatial coherence and aesthetic consistency.

The experiment also indicated that scores improved slightly for generated designs when post-processing or "refinement" by designers was applied under free prompt conditions, though this improvement was not statistically significant. This suggests that clear early-stage guidance is more effective than later refinement, consistent with the benefits observed in the deep intervention of designers under low freedom conditions.

An analysis of participants' professional backgrounds showed that non-professional participants scored higher under low freedom conditions, suggesting that deeper designer involvement ensures that generated results better meet the public's understanding and needs (Fig. 8). Conversely, participants with a professional background demonstrated better control and adjustment of AI-generated content under medium and high freedom conditions. The results of multiple linear regression analysis (Table 1) indicated that freedom level had a nearly significant effect on both aesthetic quality and alignment ($p \approx 0.05$), but the interaction with professional background was not significant, indicating that the effect of designer guidance is not dependent on the participants' professional background.

Based on these data analyses, future optimization strategies should include deep designer involvement during the initial phase to set the design framework, structured prompts in the intermediate phase to assist AI in understanding public needs, and detailed adjustments in the later stages to enhance alignment and aesthetics. Multi-phase involvement ensures the quality of generated results while maintaining the core value of designers, providing a clear collaboration pathway for efficient and creative urban planning.

## Results and discussion

### Impact of AIGC on the Role of Designers

The primary advantage of AIGC in urban planning lies in its rapid generation capability and creative potential, which allows for the swift transformation of text into images and the production of diverse design options. This significantly reduces both the time and cost required in the early design phases. However, AIGC has notable limitations in understanding complex contexts and handling comprehensive spatial strategies, particularly in overall coherence and contextual understanding. These limitations necessitate the deep involvement of designers to ensure that design outcomes meet the intended goals for spatial harmony and specific requirements.

## Value of AIGC in public participation

The use of AIGC significantly lowers the barriers to public participation in urban design, enabling non-professionals to express their design ideas through simple textual descriptions. Experimental results indicate that the quality of public-generated designs improves significantly when guided by designers. This suggests that AIGC not only aids profes-sional designers but also effectively promotes broader public engagement in urban design, facilitating a more democratized planning process.

### Role of Designers in AIGC Generation Process

Experimental data demonstrate that designer involvement significantly enhances the quality and alignment of AIGC-generated images. By providing a clear framework and structured guidance, designers help AIGC better understand the design intent, reducing inconsistencies and variability in the generated content. Furthermore, designers' creative judgment plays a key role in enhancing the aesthetic quality of generated designs, particularly in maintaining overall coherence and refining local details. Designers' expertise effectively complements the limitations of AIGC, resulting in outputs that possess higher artistic and practical value.

## Conclusion and Future Prospects

### Technical Solutions and Limitations

AIGC technology exhibits significant advantages in generating fine details, ensuring stylistic consistency, and understanding natural language within design tasks. These capabilities can be gradually optimized by enhancing training data and refining model structures. For example, improvements in AI's natural language processing can enhance its ability to understand user intentions, while advancements in image generation techniques can further improve the quality of detail and stylistic control.

However, AIGC still faces considerable limitations in areas such as overall design strategy, creative and innovative capabilities, as well as human and cultural understanding. AI lacks comprehensive spatial awareness when dealing with complex scenes and is unable to effectively integrate multidisciplinary knowledge, which limits its capacity to create truly innovative designs. Furthermore, the complexities involved in ethical considerations and cultural values are beyond the scope of current technologies. These aspects require human designers, who bring creative and holistic thinking, to effectively address such challenges. Designers' abilities in global oversight, inspiration, and cultural understanding make them irreplaceable in these areas.

### Technical Solutions and Limitations

Future research should focus on improving AI's performance in organizing spatial concepts and understanding complex design contexts, while clearly defining the roles in human-AI collaboration. AI should be employed to focus on detail-oriented and repetitive tasks, whereas designers should focus on creative and strategic design processes. Optimizing workflows for human-AI collaboration, identifying the most effective points of designer intervention, and combining AI-generated results with designers' aesthetic judgment will be crucial to enhancing both efficiency and innovation in design. Through this deep collaboration, AIGC tools can better support designers in executing complex urban planning tasks, resulting in more creative and effective solutions.